# Data-Driven Discovery of Multi-Dimensional Breakage Population Balance Equations

SUET LIN LEONG[a], FIRNAAZ AHAMED[b], YONG KUEN HO[a,c,d,1]

[a]Department of Chemical Engineering, School of Engineering, Monash University Malaysia, Jalan Lagoon Selatan, Bandar Sunway, 47500 Selangor, Malaysia

[b]School of Engineering, Faculty of Innovation and Technology, Taylor's University, 47500 Subang Jaya, Selangor, Malaysia

[c]Monash-Industry Plant Oils Research Laboratory (MIPO), Monash University Malaysia, Jalan Lagoon Selatan, Bandar Sunway, 47500 Selangor, Malaysia

[d]Centre for Net-Zero Technology, Monash University Malaysia, Jalan Lagoon Selatan, Bandar Sunway, 47500 Selangor, Malaysia

[1]To whom correspondence should be addressed.

Email: ho.yongkuen@monash.edu

**Abstract**

Multi-dimensional breakage is a ubiquitous phenomenon in natural systems, yet the systematic discovery of underlying governing equations remains a long-standing challenge. Current inverse solution techniques are restricted to one-dimensional cases and typically depend on the availability of *a priori* system knowledge, thus limiting their applicability. By leveraging advances in data-driven sparse regression techniques, we develop the Multi-Dimensional Breakage Population Balance Equation Identification (mPBE-ID) algorithm for discovering multi-dimensional breakage population balance equations (mPBEs) directly from data. Our mPBE-ID enables tractable identification of mPBEs by incorporating several key strategies, namely, a breakage-informed constrained sparse regression, targeted candidate library functions construction via insights from Dynamic Mode Decomposition (DMD), and robust handling of noisy/limited data through ensembling (bagging/bragging). Notably, we demonstrate how the DMD is indispensable for distilling dominant breakage dynamics which can then be used to facilitate the systematic inclusion of candidate library terms. We showcase the ability of the mPBE-ID to discover different forms of mPBE (including those with discontinuous stoichiometric kernels) even when tested against noisy and limited data. We anticipate that the mPBE-ID will serve as a foundational framework for future extensions to generalize the discovery of multi-dimensional PBEs for various high-dimensional particulate phenomena.

# 1. Introduction

Breakage, also known as fragmentation, is a ubiquitous and intricate phenomenon which governs many natural systems in science and engineering, e.g., in the division of biological cells, droplet breakup, granulation of drug particles in pharmaceutical manufacturing, and the formation of emulsions in colloid science. The transients of the number distribution of entities resulting from breakage can be modeled in the form of an integro-partial differential equation, known as the Population Balance Equation (PBE), [1] where the nature of breakage is embedded within the phenomenological functions, i.e., the stoichiometric and rate kernels. Crucial characteristics of fragmentation systems are often multi-dimensional, such as the length and width of high aspect ratio crystals [2-6], mass and energy of particles [7, 8], granule porosity and compositions in drug tablets [9, 10], and the chain length and degree of branching in polymers [11]. In such cases, the use of multi-dimensional breakage PBEs (mPBEs) is inevitable.

Successful implementation of population balance models (PBMs), however, hinges upon the solutions of their integro-partial differential form as well as the fidelity with which the phenomenological kernels encapsulate the intricate behaviors of particles. Since the inception of PBEs in the early 20th century, approaches to solve PBEs, including mPBEs [12-14], have been well established, but efforts to identify the phenomenological kernels in PBEs are comparatively rare. Devising breakage kernels from first principles for an unknown system is generally known to be non-trivial, hence inverse solution approaches have been pursued as alternatives to extract the phenomenology of breakage [15-27]. Although inverse solutions offer ways to uncover the underlying phenomenological functions without being committed to any specific form of the model function, they often rely on specific traits of the data such as self-similar behavior or are ill-conditioned in their formulations [28]. Moreover, the identified functions are typically non-parsimonious, and are neither *generalizable* nor *interpretable*.

Despite the increasing multi-disciplinary evidence of successful breakage PBM implementations [2, 7-11], existing inverse solution approaches for identifying the phenomenological functions of breakage PBEs are restricted to cases with one internal coordinate [15-27, 29]. Extending these existing approaches for mPBEs is expected to exacerbate the foregoing challenges encountered in one-dimensional situations. In order not to hinder scientific progress in fields where mPBEs play a crucial role, an approach that can potentially learn mPBEs from data is necessary.

In recent years, the quest to discover dynamical governing equations by leveraging the growing abundance of data has led to the emergence of 'machine scientist' [30, 31] approaches which automatically learn the governing equations for an unknown system from data. A notable approach, referred to as the sparse identification of nonlinear dynamics (SINDy), was developed by Brunton et al. [32] for the discovery of ordinary differential equations (ODEs), and was thereafter followed by the functional identification of nonlinear dynamics (PDE-FIND) by Rudy et al. [33] for discovering partial differential equations (PDEs). These approaches operate on the principle of parsimony, by using sparse regression techniques that search across a vast library of candidate functions to discover equations which are both generalizable and interpretable. These approaches, however, cannot be readily deployed to deal with PBEs with source/sink terms where integrals and partial derivatives co-exist. In the effort to re-engineer these technique for identifying PBEs, our group recently developed the Population Balance Equations Identification via Sparsity Promoting Optimization Technique (PBE-SPOT) for the discovery of one-dimensional PBEs with multiple physics, i.e., breakage, aggregation, and/or growth [34], which can either exist individually or present concurrently. Extending this technique to multiple internal coordinates to discover mPBEs, however, is highly non-trivial due to the unique set of challenges involved in navigating multi-dimensional complexities of the breakage phenomenon. As such, being able to devise a tailored data-driven algorithm based on sparse regression for identifying mPBEs is of immense value.

Leveraging the progress made in existing methods for dynamical systems discovery based on sparsity promoting techniques [32, 33, 35] and our experiences in mPBEs, here we develop an algorithmic workflow for the identification of mPBEs from particulate data. In addition to the inherent mathematical complexity of mPBEs, employing sparse regression methodologies to discover them entails two key challenges, namely the appropriate selection of candidate library functions and the ability to confront situations with noisy/limited data. The former is especially crucial in identifying mPBEs as the breakage phenomenon is an intricate process where the phenomenological stoichiometric kernels can either be continuous or discontinuous (i.e., involving a Dirac delta function). In a multi-dimensional scenario, this rapidly results in a combinatorial explosion of candidate terms, compounded by the additional challenges of implementing discontinuous functions within the library as they require additional processing such as interpolation. As with any data-driven techniques, the quality of data, i.e., noise levels and sampling rates, also critically impacts the chances of identifying the correct model. Until high-throughput high-resolution analytical measurement techniques become more commonly available/accessible, many high-dimensional particulate characterization processes are still far too labor and resource intensive for the practical purpose of data-driven estimation [36, 37].

In this work, we devise the Multi-Dimensional Breakage Population Balance Equation Identification (mPBE-ID) to remediate the foregoing challenges for the effective identification of mPBEs. We do this by embedding several key strategies in the sparse identification workflow. First, we implement a targeted library construction via first distilling key features of the breakage dynamics using the Dynamic Mode Decomposition (DMD) [38]. Being an equation-free method, DMD was initially used in flow analyses [39, 40], but have also been recently applied in other fields (e.g., neuroscience [41] and epidemiology [42]) to decompose spatiotemporal data into its static and temporal modes. Although a recent study utilized DMD to model the phenomenological kernels as black-boxes in a 1D breakage PBE [43], here we present the first application of DMD to analyze and extract key breakage dynamics. By decomposing particulate data into dominant "internal" modes (representing static modes corresponding to internal coordinates) and temporal patterns, DMD facilitates a more informed construction of the library functions for mPBE identification by extracting key breakage characteristics (e.g., continuous/discontinuous and size dependent/independent breakage). We employ Sequential Thresholded Least Squares [32] to obtain the solutions, but strategically tailor the algorithm by constraining the solution space to obtain physically realizable mPBEs. We also handle corrupt and sparse-sampled particulate data by adopting model ensembling via bagging/bragging approaches, which have proven to be promising in enhancing the robustness of sparse identification algorithms [35]. The mPBE-ID is expected to advance the field of particulate science and population balances by being the first tool capable of identifying multi-dimensional breakage PBEs directly from particulate data with the aim of yielding generalizable and interpretable particulate breakage dynamics at higher dimensions.

## 2. Methods

### 2.1 Framework for Multi-Dimensional Breakage Population Balance Equation Identification (mPBE-ID)

The multi-dimensional pure breakage population balance equation (mPBE), involving two internal coordinates and no spatial variation in the physical space, is given in the following general form:

$$\frac{\partial n(v,w,t)}{\partial t} = \underbrace{\int_w^\infty \int_v^\infty \beta(v,w,v',w')\Gamma(v',w')n(v',w',t)\,dv'\,dw'}_{\text{Birth of }(v,w)\text{ due to breakage of }(v',w')} - \underbrace{\Gamma(v,w)n(v,w,t)}_{\substack{\text{Death of }(v,w)\\ \text{due to breakage}\\ \text{of }(v,w)}} \qquad (1)$$

where $n(v,w,t)$ is the number density of particles with internal coordinates $(v,w)$ at time $t$, $\Gamma(v,w)$ is the breakage rate kernel of a particle characterized with internal coordinates $(v,w)$, and $\beta(v,w,v',w')$ is the stoichiometric kernel which represents the daughter distribution function of particles $(v,w)$ from the breakage of a parent particle characterized by internal coordinates $(v',w')$. For simplicity, the internal coordinates $(v,w)$, which in general can be any two chosen properties that characterize the particles, will be loosely referred to as sizes $v$ and $w$, each denoting a size dimension of the particle. Eq. (1) can be written compactly as:

$$\dot{n} = \Psi\big[B(\bullet), D(\bullet)\big] \qquad (2)$$

where $\dot{n} = \partial n(v,w,t)/\partial t$. Here, $\Psi$ represents a set function of terms that describe the breakage behavior of the particle population. The set function $\Psi$ is the exact solution which contains the breakage terms, i.e., the birth term $B(\bullet) = \int_w^\infty \int_v^\infty (\bullet) n(v',w',t)\,dv'\,dw'$ and the death term

$D(\bullet) = (\bullet) n(v,w,t)$. Suffice to note here that the candidate function $(\bullet)$ in $B(\bullet)$ represents the product of the breakage rate kernel and the stoichiometric kernel.

Given that we have $z$ temporal snapshots of 2D number density measurements **n**, each comprising $x \times y$ data points spanning the internal coordinate space, it becomes possible to set up a linear system of equations as follows to determine $\Psi$ from Eq. (2):

$$\dot{\mathbf{n}} = \mathbf{\Theta}\xi \tag{3}$$

where the column vector $\dot{\mathbf{n}} \in \mathbb{R}^{xyz \times 1}$ stacks the temporal partial derivatives of the number densities **n** that are numerically approximated via the finite difference method:

$$\dot{\mathbf{n}} = \begin{bmatrix} \dot{\mathbf{n}}_{11} \\ \dot{\mathbf{n}}_{12} \\ \vdots \\ \dot{\mathbf{n}}_{xy} \end{bmatrix}; \quad \dot{\mathbf{n}}_{xy} = \begin{bmatrix} \dot{n}(v_x, w_y, t_1) \\ \dot{n}(v_x, w_y, t_2) \\ \vdots \\ \dot{n}(v_x, w_y, t_z) \end{bmatrix} \tag{4}$$

Next, a library of candidate functions $\mathbf{\Theta} \in \mathbb{R}^{xyz \times c}$ is constructed, comprising two sub-libraries $\mathbf{\Theta}_B$ and $\mathbf{\Theta}_D$, i.e., $\mathbf{\Theta} = [\mathbf{\Theta}_B \ \mathbf{\Theta}_D]$, that hold potential birth terms $B(\bullet)$ and death terms $D(\bullet)$, respectively. An example of $\mathbf{\Theta}$ is shown below:

$$\mathbf{\Theta} = \left[\underbrace{B(\mathbf{1}) \ B(\mathbf{v'}) \ B(\mathbf{w'}) \ B(\mathbf{v'w'}) \ B(\mathbf{1/(v'w')}) \ \cdots}_{\mathbf{\Theta}_B} \ \underbrace{D(\mathbf{1}) \ D(\mathbf{v}) \ D(\mathbf{w}) \ D(\mathbf{vw}) \ \cdots}_{\mathbf{\Theta}_D}\right] \tag{5}$$

For ease of representation, we employ notations such as **v'w'** or **vw** to denote column vectors of which the function $v'w'$ or $vw$ is being embedded into the birth or death term. To provide a clearer illustration of the notations, example columns of $\mathbf{\Theta}$ evaluated for the entire data set are shown below for $B(\mathbf{v'w'})$ and $D(\mathbf{vw})$:

$$B(\mathbf{v'w'}) = \begin{bmatrix} \int_{w_1}^{\infty}\int_{v_1}^{\infty} v'w'n(v',w',t_1)dv'dw' \\ \int_{w_1}^{\infty}\int_{v_1}^{\infty} v'w'n(v',w',t_2)dv'dw' \\ \vdots \\ \int_{w_y}^{\infty}\int_{v_x}^{\infty} v'w'n(v',w',t_{z-1})dv'dw' \\ \int_{w_y}^{\infty}\int_{v_x}^{\infty} v'w'n(v',w',t_z)dv'dw' \end{bmatrix}; \quad D(\mathbf{vw}) = \begin{bmatrix} v_1 w_1 n(v_1,w_1,t_1) \\ v_1 w_1 n(v_1,w_1,t_2) \\ \vdots \\ v_x w_y n(v_x,w_y,t_{z-1}) \\ v_x w_y n(v_x,w_y,t_z) \end{bmatrix} \quad (6)$$

All integrals can be approximated numerically such as via the trapezoidal rule or Simpson's rule. The functions embedded in the candidate terms are not limited to polynomial and rational functions listed in Eq. (5). For example, more complex stoichiometric kernels, such as Dirac delta functions and Heaviside functions can be incorporated in $\Theta_B$ if necessary, depending on the known nature of the system. Certainly, incorporating relevant functions into the library is highly pivotal for sparse and accurate identification of mPBEs while reducing computational expense. Our method for constructing candidate library functions through data-driven insights via DMD is elucidated in Section 2.3.

In formulating Eq. (3), mPBE-ID aims to infer $\Psi$ and derive a parsimonious mPBE, prioritizing interpretability by selecting a sparse subset of columns in $\Theta$. The column vector $\xi$ in Eq. (3) represents a sparse vector where its non-zero entries determine the active columns in $\Theta$. As a result, the identified mPBE is represented as a weighted linear combination of the active columns of $\Theta$, with the weights given by the column vector $\xi$. Based on breakage-informed knowledge on the architectural form of mPBE, which includes a positive birth term and a negative death term to satisfy the law of mass conservation, the mPBE-ID solves Eq. (3) via a constrained sparse regression protocol, as detailed in Section 2.4.

## 2.2 Conceptual Description of 2D Breakage Continuity Scenarios

Although 2D breakages need not necessarily be geometrically driven, for simplicity we illustrate the concept of continuous/discontinuous breakages using 2D geometrical (rectangular) examples as illustrated in Fig. 1. In these scenarios, the daughter probability distribution can exhibit three distinct forms of continuity across the internal coordinates: continuous, semi-continuous, and discontinuous. As shown in Fig. 1a, particles may break at any random position within the rectangle, resulting in a continuous distribution of daughter particle sizes in both dimensions [12, 44]. Conversely, a breakage scenario could occur where one dimension breaks while the other remains intact as illustrated in Fig. 1b [44]. This results in a semi-continuous probability distribution where one dimension remains continuous while the other is discrete. Additionally, particles may break at specific locations (i.e., at the midpoint), leading to the formation of equal-sized fragments [45], as depicted in Fig. 1c, and hence the stoichiometric kernel is discontinuous in both size dimensions.

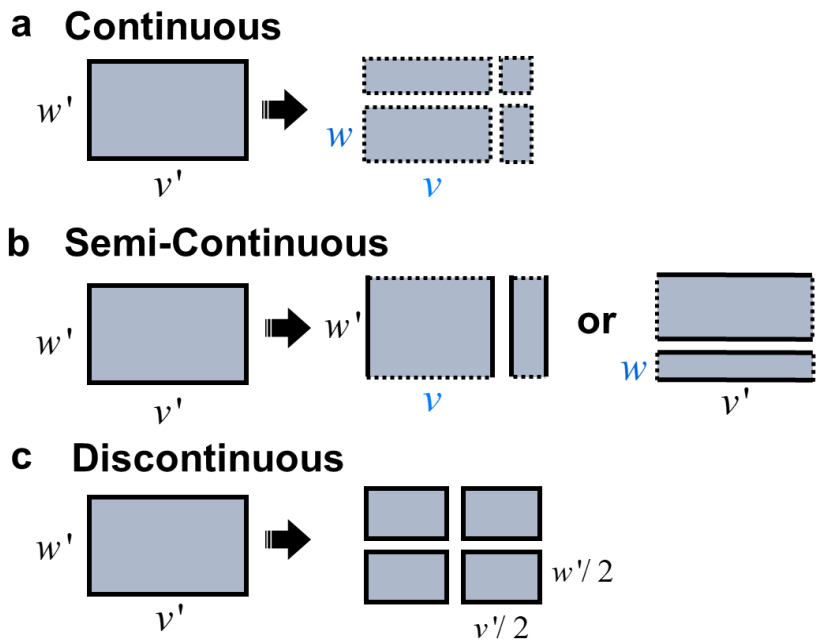

Figure 1 Example conceptual illustrations of 2D breakage of a rectangular particle with stoichiometric kernels, $\beta(v,w,v',w')$, or daughter distributions that are: (a) continuous, (b)

semi-continuous, and (c) discontinuous. Here, *v* and *w* represent the length and width of the daughter particles, while *v'* and *w'* denote the corresponding dimensions of the parent particle. The edges of the daughter particles are dashed if they follow a continuous distribution, whereas a solid edge indicates a fixed size, e.g., equal to the parent size (b) or half the parent size (c).

## 2.3 Construction of Library of Candidate Functions via Insights from Dynamic Mode Decomposition (DMD)

Fundamental to the identification of mPBEs by mPBE-ID is the construction of a well-suited library of candidate terms $\Theta$ (Eq. (5)). In general, we recommend including a broad range of basic mathematical terms, such as polynomial and rational functions, as they effectively represent common continuous breakage forms (e.g., random breakage) [7, 46, 47]. When building the birth term library $\Theta_B$, the need to incorporate possible forms of stoichiometric kernels with different continuity scenarios may arise. Notably, discontinuous breakage functions, specifically the Dirac delta functions, may require additional treatment to the data (e.g., data interpolation) as the birth integrals necessitate evaluation at specific data sampling sizes to populate $\Theta_B$. When no information about the potential physics is available *a priori*, the potential coupling of the stoichiometric kernels with different continuity scenarios and rate kernels within the birth term (Eq. (1)) rapidly leads to a combinatorial explosion of library terms in $\Theta_B$. Hence, being able to identify key breakage patterns in the probability distribution functions before library construction is crucial for narrowing the selection of candidate breakage kernels, leading to a more targeted and efficient library.

To achieve this, we adopt the Dynamic Mode Decomposition (DMD) to extract key patterns in the temporal and particle internal coordinate space from the data matrix **n**. Briefly, from the given data matrix **n**, it is possible to arrange $z$ snapshots into two $xy \times (z-1)$ data matrices as shown below:

$$\mathbf{X} = \begin{bmatrix} | & | & & | \\ \mathbf{n}_1 & \mathbf{n}_2 & \cdots & \mathbf{n}_{z-1} \\ | & | & & | \end{bmatrix}; \quad \mathbf{X'} = \begin{bmatrix} | & | & & | \\ \mathbf{n}_2 & \mathbf{n}_3 & \cdots & \mathbf{n}_z \\ | & | & & | \end{bmatrix} \quad (7)$$

The DMD computes the leading eigendecomposition of the best-fit linear operator $\mathbf{A}$ that relates the data matrices $\mathbf{X'} \approx \mathbf{AX}$:

$$\mathbf{A} = \mathbf{X'X}^{\dagger} \quad (8)$$

where $\dagger$ is the Moore-Penrose pseudoinverse. Each eigenvalue-eigenvector pair of $\mathbf{A}$ represents the temporal frequency and the corresponding DMD mode (or internal modes in the context of distributed particulate data). The pseudoinverse $\mathbf{X}^{\dagger}$ can be approximated via singular value decomposition (SVD) of the matrix $\mathbf{X}$ with a truncation $s$. For algorithmic details on computing the lower rank approximation of DMD modes and the corresponding temporal frequencies, readers are referred elsewhere [38-40].

The original data can be reconstructed from the DMD modes and eigenvalues through their linear combination and the end result is given via the following expression [48]:

$$\hat{\mathbf{n}}(t) = \sum_{j=1}^{s} \mathbf{\Phi}_j \exp(\Omega_j t) \mathbf{b}_j \quad (9)$$

where $\hat{\mathbf{n}}(t)$ is a rank-$s$ approximation of the number density $\mathbf{n}(t)$ at time $t$, $\mathbf{\Phi}_j$ is the column vector of the $j$-th DMD mode, $\Omega_j = \ln(\Lambda_j)/\Delta t$ is the continuous-time eigenvalue associated with the $j$-th DMD mode, $\Lambda_j$ is the corresponding $j$-th discrete-time eigenvalue, and $\mathbf{b}$ is the column vector of mode amplitudes to match the initial condition. In particular, the temporal dynamics encoded in the eigenvalues offer valuable insights. The magnitude of discrete-time eigenvalue $\Lambda_j$ reflects the spectral radius, $r_j = |\Lambda_j|$, and indicates whether the corresponding mode grows or decays over time, while its argument (or phase angle), $\theta_j = \arg(\Lambda_j)$, represents the oscillation frequency when plotted on the complex unit circle. Specifically, when $r_j < 1$, the corresponding DMD mode decays over time, whereas $r_j > 1$ indicates growth (Eq. (9)). Moreover, for $r_j < 1$, smaller $r_j$ values yield a more negative $\ln(\Lambda_j)$, which in turn makes the real part of $\Omega_j$ more negative, resulting in faster decay of $\exp(\Omega_j t)$ in Eq. (9). The same reasoning applies for growing modes. These modes reflect rapid fragmentation events within the breakage system. In contrast, modes with $r_j$ close to but less than unity (i.e., near the unit circle in the complex plane) decay slowly, capturing long-timescale behaviors that may correspond to slower breakage dynamics. Thus, the spectral radii of the DMD modes serve as a proxy for distinguishing the size dependence of the breakage rate. This distinction is crucial for determining the appropriate forms of candidate functions to include in the death term library $\mathbf{\Theta}_D$ of which only the rate kernel is unknown (Eq. (1)). As for the DMD modes, we expect distinct distribution patterns to manifest in fundamentally different systems, potentially revealing the dominant internal coordinate distributions. This enables the identification of key breakage patterns, especially the continuity of breakage, as described in Section 2.1, which allows the inference of possible structural forms of the stoichiometric kernels for inclusion in $\mathbf{\Theta}_B$. The possible structural forms of the stoichiometric and rate kernels divulged from DMD would subsequently allow a systematic inclusion of candidate birth terms in $\mathbf{\Theta}_B$ due to their multiplicative nature within the birth term (Eq. (1)).

Fig. 2 shows a conceptual illustration of the model predicted without DMD vs. with DMD. Without prior information regarding the breakage characteristics, a natural basis is to employ a library with continuous terms as the evaluation of the library candidate terms does not require additional data preprocessing. However, DMD reveals discontinuities in the DMD modes paired with a size-independent breakage rate from the eigenvalue spectrum, which substantially advises the reduction in the number of library terms. This not only improves the accuracy of number density prediction but more importantly facilitates a more efficient sparse and interpretable identification of mPBEs. Details on how DMD mode patterns distinguish breakage systems and how the eigenvalue spectrum reveals rate dependency on internal coordinates, and their use in curating the library matrix are provided in Section 3.3.

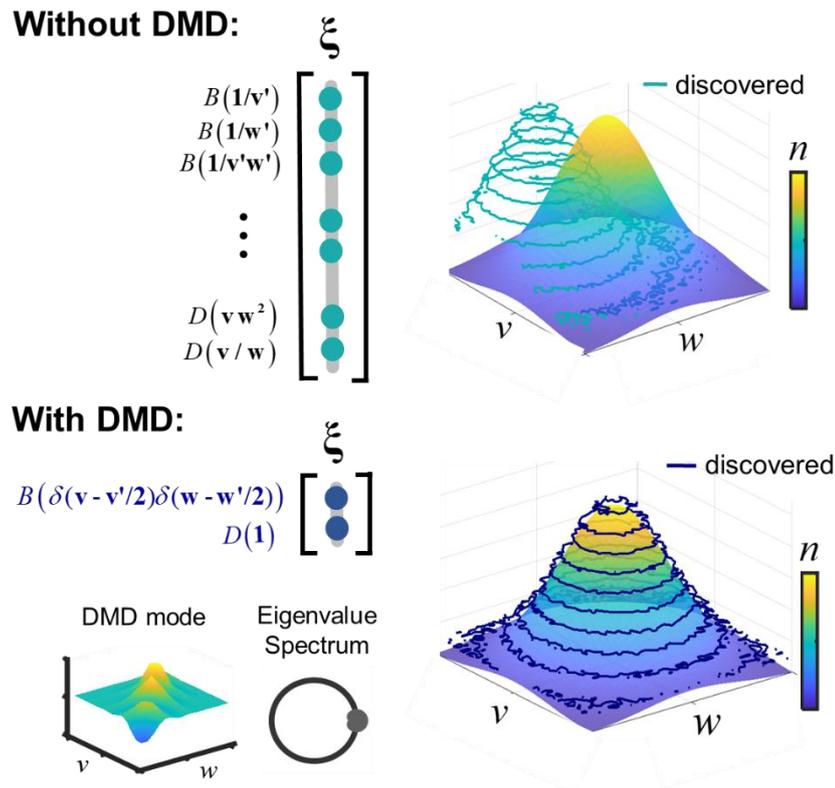

Figure 2 Comparison of model identification without and with DMD using an example with a 2D breakage scenario that yields equal-sized fragments with a size-independent rate. When no information about the physics is available *a priori*, identification without DMD employs a library containing continuous polynomial and rational functions which are natural guesses for particulate breakage. However, DMD modes inform the need to incorporate Dirac delta functions to capture discontinuous breakage types, while the eigenvalues indicate a size-independent breakage. In the spirit of promoting a minimalist and efficient library construction, leveraging DMD to shed such insights from data proves promising to identify the correct breakage dynamics.

## 2.4 Constrained Ensemble Sparse Identification of mPBEs

Upon setting up the library $\boldsymbol{\Theta}$ via insights from DMD, we solve for the sparse vector $\boldsymbol{\xi}$ in Eq. (3) for mPBE identification. In this work, we employ the efficient Sequential Thresholded Least Squares (STLS) [32], which iteratively performs a least squares fit and sets elements in $\boldsymbol{\xi}$ below a threshold $\lambda$ to zero until convergence. However, the inherent structure of mPBEs, governed by the law of mass conservation, necessitates a net positive birth term and a net negative death term to ensure its physical realizability. To enhance the chances of obtaining realizable mPBEs, instead of an unconstrained least-squares in the original STLS, we constrain the solution space as such:

$$\min_{\boldsymbol{\xi}} \|\dot{\mathbf{n}} - \boldsymbol{\Theta}\boldsymbol{\xi}\|_2^2 \quad s.t. \quad \begin{bmatrix} -\boldsymbol{\Theta}_B & \mathbf{0} \\ \mathbf{0} & \boldsymbol{\Theta}_D \end{bmatrix} \boldsymbol{\xi} \leq \varepsilon \begin{bmatrix} -1 \\ -1 \end{bmatrix}; \quad \varepsilon > 0 \tag{10}$$

where $\varepsilon$ is a small, positive non-zero value (e.g., $\varepsilon = 1e-4$). The inequality constraints in Eq. (10) are designed to ensure that the linear combination of birth terms obtained from the least-squares solution remains overall positive, and similarly, that of the death terms remains overall negative. Note that the constrained least squares problem in Eq. (10) is only feasible when the constraint matrix $[-\boldsymbol{\Theta}_B \ \mathbf{0}; \ \mathbf{0} \ \boldsymbol{\Theta}_D]$ is free from rows with all zero entries. Ideally, the inequality constraints should be enforced across all rows of $\boldsymbol{\Theta}$ to ensure full physical consistency. Nevertheless, to avoid prohibitive computational cost, a relaxation of Eq. (10) can be invoked by imposing the constraints on a selected subset of $\boldsymbol{\Theta}$ rows to effectively steer the solution toward physically realizable PBE regions. We call the proposed algorithm the Constrained Breakage-Informed Sequential Thresholded Least Squares (cb-STLS).

To improve the robustness of the algorithm against noise and scarcity in data, we employ data bootstrapping and model ensembling strategies. We do this by generating an ensemble of models via deploying cb-STLS over multiple bootstrap data samples, i.e., resampling the original dataset with replacement, for a specific value of $\lambda$, after which the ensemble of models are then aggregated by either taking the mean (bagging) or median (bragging) of the identified coefficients [35]. The coefficients in the aggregated model are then thresholded based on both inclusion probability and coefficient of variation. The inclusion probability quantifies the frequency in which a coefficient is non-zero across the ensemble while the coefficient of variation (i.e., ratio of standard deviation to mean) reflects the variability in coefficient magnitude across the ensemble. Coefficients are retained only if they exceed a minimum inclusion probability and exhibit sufficiently low variability across bootstrap replicates. The entire bootstrapping and model ensembling process are repeated for different values of $\lambda$ to generate a pool of candidate models with varying sparsity, where the optimal model is then selected via a screening cost function elucidated in the next section.

**2.5 Model Selection**

From the pool of candidate models with varying complexity, we seek to identify the optimal model that balances accuracy, sparsity, and importantly, of which comprises a valid mPBE structure. While the constrained least squares algorithm in Eq. (10) ensures the presence of both birth and death terms in the initial solution, subsequent thresholding, both via STLS and the final aggregated model obtained through ensembling, may eliminate one of these terms, leading to an invalid mPBE (i.e., lacking a birth or death term). As such, to facilitate model selection, we impose a penalty term for penalizing invalid mPBEs via the third term in the following cost function, along with penalties for model error and sparsity:

$$\boldsymbol{\xi} = \underset{\boldsymbol{\xi}'}{\operatorname{argmin}} \; \lambda_1 \ln \| \dot{\mathbf{n}} - \boldsymbol{\Theta}\boldsymbol{\xi}' \|_2^2 + \lambda_2 \|\boldsymbol{\xi}'\|_0 + \lambda_3 \mathrm{I}\left( \|\boldsymbol{\xi}'_B\|_0 \cdot \|\boldsymbol{\xi}'_D\|_0 = 0 \right) \qquad (11)$$

where $\lambda_i$ ($i$ = 1, 2, 3) are the individual weights in the cost function, I($\bullet$) is the indicator function, and $\boldsymbol{\xi}'_B$ as well as $\boldsymbol{\xi}'_D$ are the identified birth and death term coefficients, respectively.

Note that models comprising all zero-valued coefficients are excluded from the model selection process. The first two terms of Eq. (11) effectively balance the least squares error and the sparsity of the solution vector $\xi'$. The third term is what we call the mPBE regularizer, which imposes penalties on solutions that violate the structural integrity of mPBEs. Since the architecture of mPBEs necessitates the coexistence of both birth and death terms, the mPBE regularizer is crucial to penalize against structurally invalid solutions. The regularizer operates through an indicator function $I(\cdot)$ that assumes unity when either birth or death term is absent, and zero when both terms are present.

Fig. 3 summarizes our approach, while Fig. 4 gives the detailed implementation protocol for mPBE-ID.

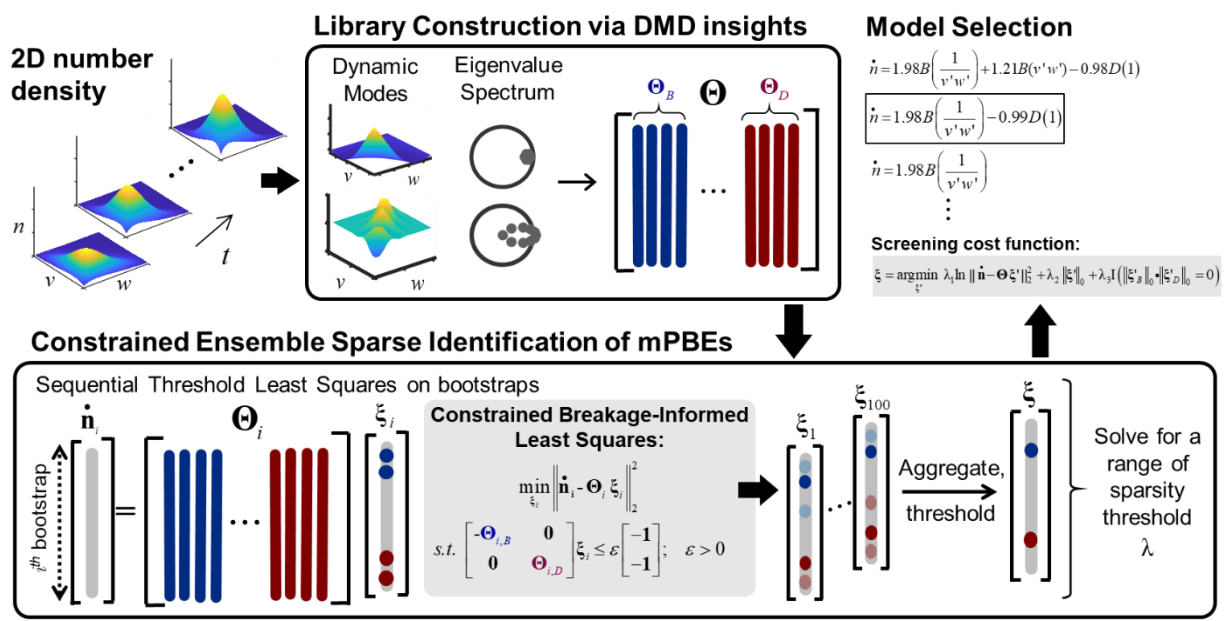

Figure 3 Overview of the Multi-Dimensional Breakage Population Balance Equation Identification (mPBE-ID). With time-series measurements of 2D number densities, the Dynamic Mode Decomposition (DMD) is first deployed to distill the dominant internal modes and eigenvalue spectrum for targeted library construction. The mPBE-ID then solves the identification problem via Constrained Breakage-Informed Sequential Thresholded Least Squares (cb-STLS) on $i$ data bootstraps (sampling with replacement) for a specific threshold value $\lambda$. The ensemble of $i$ models generated from bootstraps is aggregated by taking the mean/median of coefficient values, with thresholding on inclusion probabilities and coefficients of variation applied to obtain a sparse vector $\xi$. With a range of threshold values used in cb-STLS, a pool of models is generated which are screened using a customized cost function that balances accuracy, parsimony, and structural validity of the mPBEs.

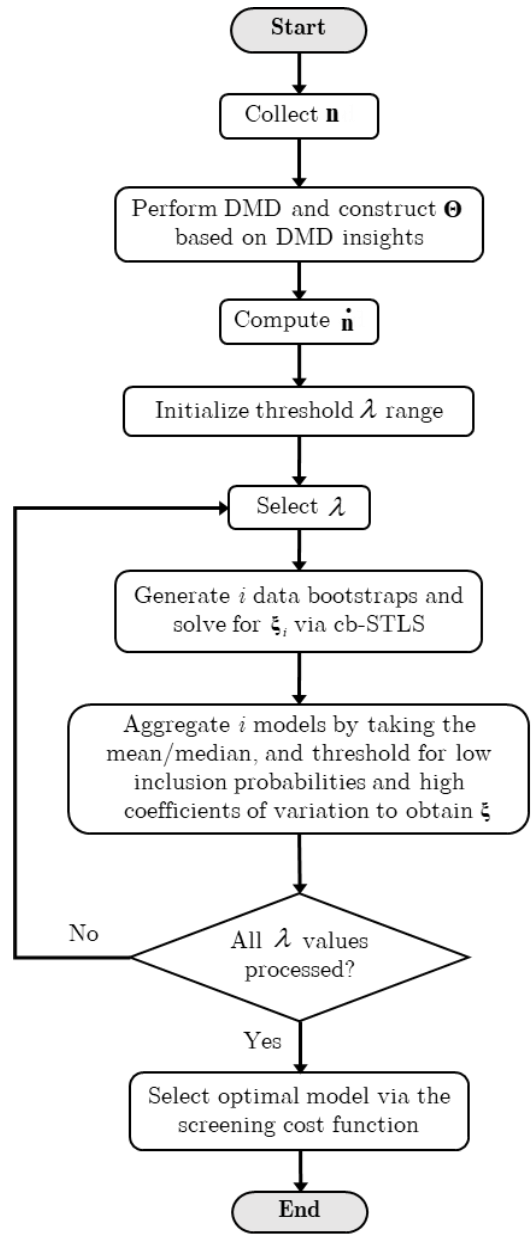

Figure 4 Algorithmic implementation flowchart of Multi-Dimensional Breakage Population Balance Equation Identification (mPBE-ID). Here, DMD refers to the Dynamic Mode Decomposition and cb-STLS the Constrained Breakage-Informed Sequential Threshold Least Squares (see Section 2.4).

## 3. Results and Discussion

### 3.1 Preamble: Description of Case Studies and Data Generation

In this work, we test the performance of the mPBE-ID on transient number density data from several known 2D breakage scenarios in the literature, presented in Table 1. Here, the cases can be divided into three main categories: (i) continuous breakage (Cases 1-4) [12, 44], (ii) semi-continuous breakage (Case 5) [44], and (iii) discontinuous breakage (Case 6) [45]. The first category involves a random breakage process, where particles fragment into a continuous spectrum of sizes, occurring in a quaternary (Cases 1-2) or binary fashion (Cases 3-4). The second category represents the binary fragmentation of a geometrical object where one size dimension breaks, while the other remains unchanged, a pattern commonly observed in the breakage of high aspect ratio crystals [2, 44]. The final case describes discrete fragmentation events, where the object breaks into four equally sized fragments in both the $v$ and $w$ dimensions.

Table 1 Cases employed for evaluating the performance of mPBE-ID. Here, $\beta(v,w,v',w')$ is the stoichiometric kernel which gives the distribution of particle of size $(v,w)$ due to the breakage of a parent particle of size $(v',w')$ and $\Gamma(v,w)$ is the breakage rate kernel for a particle of size $(v,w)$. In the mPBEs, $B(\bullet) = \int_w^\infty \int_v^\infty (\bullet) n(v',w',t) dv' dw'$ and $D(\bullet) = (\bullet) n(v,w,t)$ are the birth and death terms (Eq. (1)), respectively.

| Case | $\beta(v,w,v',w')$ | $\Gamma(v,w)$ | mPBE form |
|---|---|---|---|
| Continuous breakage | | | |
| 1 | $\dfrac{4}{v'w'}$ | 1 | $\dot{n} = 4B\left(\dfrac{1}{v'w'}\right) - D(1)$ |
| 2 | | $vw$ | $\dot{n} = 4B(1) - D(vw)$ |
| 3 | $\dfrac{2}{v'w'}$ | 1 | $\dot{n} = 2B\left(\dfrac{1}{v'w'}\right) - D(1)$ |
| 4 | | $v+w$ | $\dot{n} = 2B\left(\dfrac{1}{v'}\right) + 2B\left(\dfrac{1}{w'}\right) - D(v) - D(w)$ |

| | Semi-Continuous breakage | | |
|---|---|---|---|
| 5 | $\dfrac{v'\delta(v-v')+w'\delta(w-w')}{v'w'}$ | $vw$ | $\dot{n}=B\left(v'\delta(v-v')\right)+B\left(w'\delta(w-w')\right)-D(vw)$ |
| | Discontinuous breakage | | |
| 6 | $4\delta(v-v'/2)\delta(w-w'/2)$ | $\dfrac{1}{4}$ | $\dot{n}=B\left(\delta(v-v'/2)\delta(w-w'/2)\right)-\dfrac{1}{4}D(1)$ |

To generate time-series number density data, we employed analytical solutions where available (Case 1 [44] and Case 6 [45]) and used the highly accurate Fixed Pivot Technique [13, 49] for cases without an analytical solution (Cases 2-5). For Cases 1-5, we employed an internal coordinate domain of $[10^{-1},5]\times[10^{-1},5]$ meshed with 25 × 25 geometrically spaced grid points and a time domain of $[0,5]$. Data was generated using a monodisperse initial distribution, $n(v,w,0)=N_0\,\delta(v-v_{\max})\delta(w-w_{\max})$, where $N_0$ is the initial number of particles per unit volume. For Case 6, due to the nature of breakage, we employed a polydisperse initial distribution, $n(v,w,0)=N_0\left(16vw/\left(v_0^2 w_0^2\right)\right)\exp\left[-2\left(v/v_0+w/w_0\right)\right]$ [14, 50], to generate more diverse daughter particle sizes to populate the entire 2D mesh for more data sampling points. Here, $v_0$ and $w_0$ are the initial sizes and are both set to unity and $N_0=1$ for all cases. For Case 6, due to the rapid dynamics, we use a domain of $[10^{-4},5]\times[10^{-4},5]$ meshed with 15 × 15 grid points and a time domain of $[0,1]$. Since the daughter particles are half the size of the parent particles in Case 6, it is convenient (for the sake of testing the performance of mPBE-ID) to implement a mesh with a geometric ratio of two for the integration of Dirac delta functions in the library. Nevertheless, in situations where the experimentalist collects data with a different resolution of the internal coordinates, the number densities at the discrete sizes can be obtained with the help of interpolation techniques. As for the temporal sampling frequencies, an analysis of their effect on mPBE-ID performance is elucidated in Section 3.5.

## 3.2 Preamble: Algorithmic Setup for mPBE-ID and Performance Metrics

All computations in this work were carried out using MATLAB® R2024a on a workstation equipped with an Intel® Core™ i5-1135G7 CPU, a clock speed of 2.40 GHz, and 16 GB of installed memory (RAM). Throughout all case studies, the sparse solution of Eq. (3) was solved using sparsity values λ ranging from 0.1 to 1, with increments of 0.1. Note that this range is suitable for the data used in this study. As a general guide, one may first perform least-squares (without thresholding) to identify the maximum values of the weights and impose the range of λ accordingly. To prevent infeasibility in enforcing inequality constraints, rows where both the birth and death components had small magnitudes (sum of absolute values < 0.1) were excluded from $\Theta$. In addition, the value of $\varepsilon$ was set to 1e-4 in cb-STLS (Eq. (10)). When solving the constrained least squares in cb-STLS, we employed the "lsqlin" sub-routine, with both the step tolerance and constraint tolerance set at 1e-4, and the maximum iteration limit at 1000. For bootstrap implementation, the "bootstrp" sub-routine was employed with a total of 100 bootstraps performed for each sparsity threshold λ, and parallel execution was enabled to accelerate computation. When aggregating the ensemble, model coefficients with inclusion probabilities < 65% or coefficient of variation > 1 were set to zero. While these thresholds were fixed in this study, they are not meant to be definitive and reflect possible trade-off between model parsimony and fitting performance. In practice, these values can be varied systematically across a broad range during model screening, and refined iteratively based on model validation results to arrive at a satisfactory model.

When screening candidate models using the cost function in Eq. (11), penalty weights were set to $\lambda_1 = 1$, $\lambda_2 = 10$, and $\lambda_3 = 1000$, with higher weights assigned for the third penalty term to prioritize the attainment of physically valid mPBEs. In general, the third penalty term (penalizing invalid mPBEs) should be assigned a weight significantly larger than those of the first two terms, which correspond to model error and parsimony. Nonetheless, our framework is designed to screen models over a broad hyperparameter space, and when computational resources allow, users can perform iterative validation to identify the optimal model. To evaluate the performance of the mPBE-ID for PBE identification, both the success rate and error in the model coefficients were quantified. The success rate is the proportion of correctly identified functional terms in the model from the library divided by the total number of library terms. However, the success rate does not account for the departure in the values of model coefficients with respect to the true values. As such, we use the model coefficient error $E_c$ to quantify the error between the model coefficients of the identified terms $\xi$ and the true coefficients $\hat{\xi}$, following Fasel et al. [35]:

$$E_c = \frac{\|\hat{\xi} - \xi\|_2}{\|\hat{\xi}\|_2} \tag{12}$$

### 3.3 Insights on Particle Breakage Dynamics from DMD

The dynamic mode decomposition of the 2D number density for continuous breakage cases is presented in Fig. 5, and for semi-continuous and discontinuous cases in Fig. 6, illustrating the dominant internal modes (or DMD modes) and their temporal dynamics via the discrete eigenvalue spectra. As alluded to in Section 2.3, each DMD mode corresponds to an eigenvalue which represents the time dynamics, with discrete eigenvalues inside the unit circle representing decaying dynamics ($|\Lambda_j| < 1$), while those outside the unit circle corresponding to growing dynamics ($|\Lambda_j| > 1$). The DMD modes are numbered and paired with their associated eigenvalues. When interpreting the DMD analyses, note that DMD modes with negative magnitudes, when paired with negative temporal dynamics (Fig. A.1 in Appendix), will result in a positive contribution to the overall dynamics. For all cases, the data used consists of 25 evenly spaced timepoints sampled over the temporal domain [0, 5] for Cases 1-5 and [0, 1] for

Case 6. Here, the data used is noise-free (i.e., clean), and the interpretation of breakage dynamics from DMD remains unaffected under the addition of a small level of noise ($\leq 5$ %). The degree of truncation $s$ used in the SVD computation during DMD execution is 10, retaining > 99% of the total energy in all cases. It is also noteworthy that oscillatory modes appear in complex-conjugate pairs and, for brevity, only one DMD mode of the pair is plotted. Thus, six DMD modes are shown (out of ten) due to four oscillatory modes in Cases 1-5, while five modes are shown for Case 6 due to five oscillatory modes. These duplicated modes are also reflected in the eigenvalue spectra (Figs. 5-6) as eigenvalues with identical real parts but with imaginary parts that are equal in magnitude and opposite in sign.

### 3.3.1 Continuous Breakage

Fig. 5 illustrates the DMD of the distributions associated with continuous forms of breakage (Cases 1-4) stemming from a monodisperse initial condition. Note that the insights from DMD with a polydisperse initial condition are similar and hence are not shown. From Fig. 5, the DMD eigenvalues reveal distinct temporal behaviors across different modes. Mode 1 consistently exhibits growing dynamics in most cases (Cases 1, 2, and 4), while modes such as Modes 5-6 demonstrate decaying behavior across all scenarios. This aligns well with physical expectations: modes with higher magnitudes at smaller particle sizes (e.g., Mode 1 of Cases 1, 2, and 4) reflect the accumulation of fragments over time, while modes peaking at larger sizes (e.g., Mode 5 of Cases 1 and 4, Mode 6 of Cases 2 and 3) capture the progressive disappearance of larger particles due to breakage over time (Fig. 5), which exemplifies the fundamental behavior of breakage systems. In Case 1, most of the eigenvalues lie outside the unit circle unlike other cases, which indicate that the modes are primarily growing over time and are unstable. This behavior is expected, as the fragmentation mechanism, involving quaternary random breakage combined with a size-independent rate, causes a rapid rise in smaller fragments compared to other cases, within the same time interval studied.

From Fig. 5, the DMD modes clearly reveal that the distribution spans the entire range of the internal coordinate domain for Cases 1-4, reflecting a continuous probability distribution in the stoichiometric kernel. As a result, polynomial and rational functions are suitable candidates for the daughter distribution function in the birth term library as they reflect a daughter distribution function that is continuous. By further inspecting the DMD eigenvalues, we observe distinct distribution patterns: in cases with size-independent breakage rates (Cases 1, 3), the eigenvalues cluster about the same spectral radius, reflecting uniform decay/growth rates across all internal modes. In contrast, for size-dependent breakage (Cases 2, 4), the eigenvalues are more widely distributed across the unit circle, indicating that different internal modes, associated with peaks at different particle sizes, evolve at varying frequencies. This observation aligns with the theoretical interpretation alluded to earlier (Section 2.3), where the spectral radius serves as a tool to identify the size dependence of the breakage rate.

The insights drawn from DMD regarding the size-dependency in the breakage rate from the eigenvalues coupled with breakage characteristics revealed from the DMD modes allow for systematic inclusion of library candidate terms. For the death term library $\mathbf{\Theta}_D$, a permutation of size-dependent terms should be incorporated, which generally includes basic polynomial and rational functions if the eigenvalues suggest size-dependency behavior. Otherwise, it reduces to $D(1)$. As for the birth term library $\mathbf{\Theta}_B$, a range of polynomial and rational functions should be included as both the stoichiometric and rate kernels encompass similar mathematical structures.

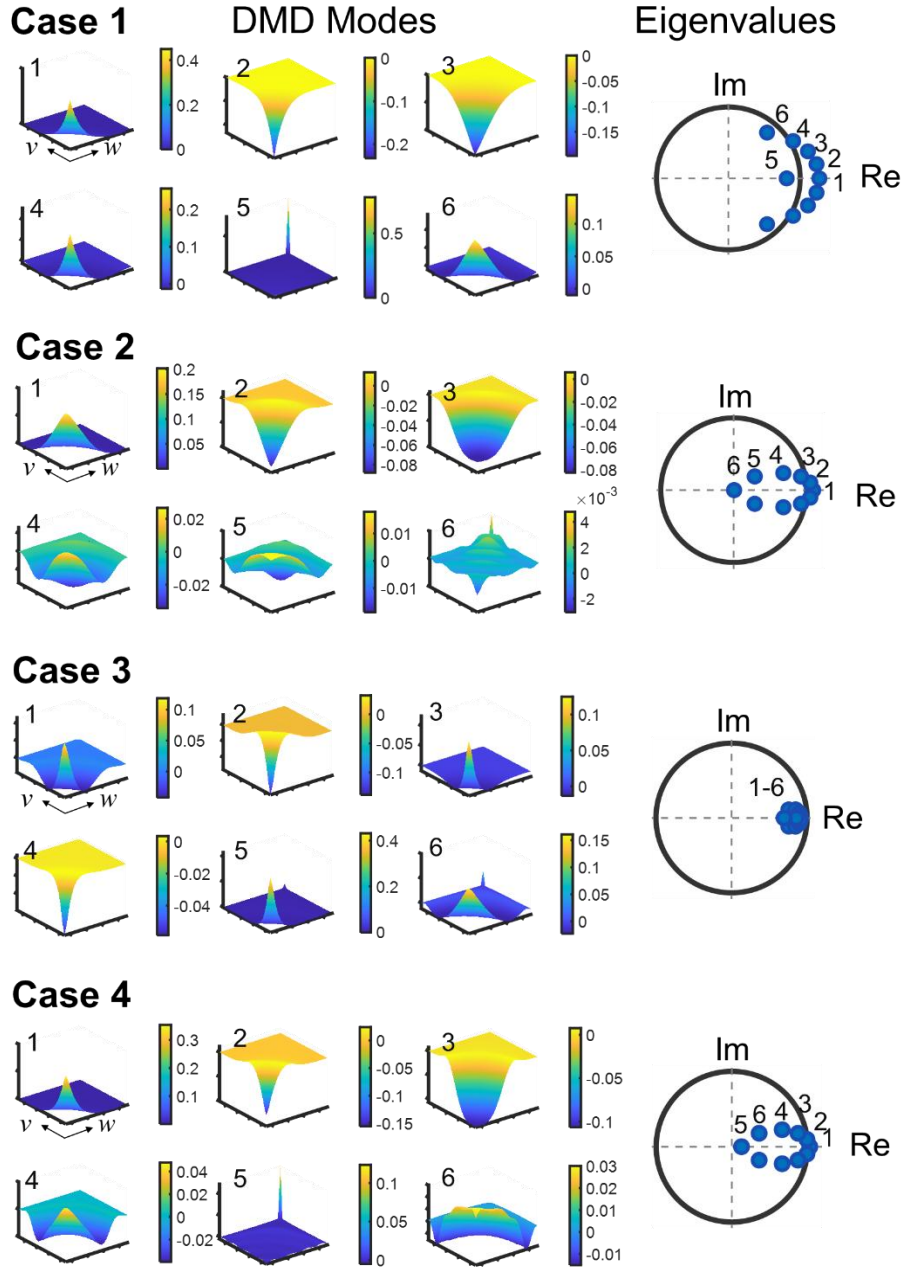

Figure 5 DMD modes $\Phi_j$ and discrete eigenvalues $\Lambda_j$ of 2D continuous breakages (Cases 1-4). Clean data were used, sampled at 25 timepoints over the time domain [0, 5], with distributions derived from a monodisperse initial condition. The numbering of the DMD modes and eigenvalues corresponds within each case. The corresponding continuous-time dynamics of each DMD mode are provided in the Appendix. The degree of truncation used in the SVD is $s = 10$, capturing > 99% of the total energy. Oscillatory modes appear in complex-conjugate pairs and only one of each pair is shown for brevity. The colorbar represents the magnitude of mode contribution to the number density distribution.

### 3.3.2 Semi-Continuous Breakage

The DMD analyses of a semi-continuous breakage (Case 5) are shown in Fig. 6a resulting from both a monodisperse and a polydisperse initial distribution. Notably, a distinct pattern emerges within the internal coordinate domain when discontinuity arises as compared to those in continuous breakages (Fig. 5). With a monodisperse initial distribution, the DMD modes reveal a cascading pattern of number density distributions along the edge of the internal coordinate domain, more clearly observed in Modes 3-6 in the 3D view as indicated by the arrows in Fig. 6a(i). This reflects a directionally selective fragmentation, where breakage occurs along a preferred axis while the other dimension remains largely unchanged, thereby suggesting an overall semi-continuous breakage behavior. However, the pattern is less distinctive when a polydisperse initial condition is employed (Fig. 6a(ii)), though it remains discernible through an observable anti-diagonal peak (extending from top left to bottom right of domain, more evident in the 2D view) that propagates across the domain in all modes. The patterns observed from DMD modes are particularly impressive as such discontinuities are not easily discernible from the raw data. These mode patterns evince that the stoichiometric kernel is likely composed of a linear combination of Dirac delta functions in both dimensions, i.e., $\delta(v-v')$ and $\delta(w-w')$, to maintain uniformity along one dimension as the other undergoes breakage. Nevertheless, care must be exercised, as anti-diagonal features in the DMD modes can also be indicative of fully continuous breakage, as observed in Mode 6 of Case 2 (see Fig. 5). Thus, it is advisable to always analyze the scenario via both 2D/3D plots of DMD modes for a more holistic interpretation, or to consider performing experiments with monodisperse initial distribution where allowable to amplify the signature behavior. Alternatively, it may be prudent to include both semi-continuous and continuous terms in the candidate library when such patterns arise, albeit at the expense of increased computational cost due to the expanded library.

Further, it is evident that in all cases, breakage occurs symmetrically across both internal coordinates, indicating a uniform daughter distribution function and rate of breakage in both directions. If these factors were asymmetric, the DMD modes would appear skewed (not shown here), which would indicate a directional bias in breakage, such as a higher probability of breakage along the length compared to the width in high aspect ratio crystals [44]. As for the eigenvalues, their wide spread distribution within the spectra as seen in Fig. 6a reflects a size-dependent rate which is similarly observed in Cases 2 and 4 with a size-dependent rate kernel (Fig. 5), and thus polynomial and rational functions would similarly be appropriate to populate $\Theta_D$. This, together with the possible stoichiometric kernel structure, would result in a permutation of $v'^p w'^q \cdot \delta(v-v')$ and $v'^p w'^q \cdot \delta(w-w')$ for $\Theta_B$.

## a  Semi-Continuous Breakage

### i) Case 5 (with monodisperse IC)

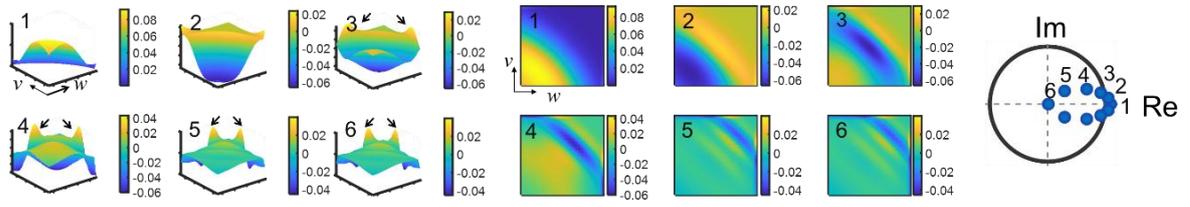

### ii) Case 5 (with polydisperse IC)

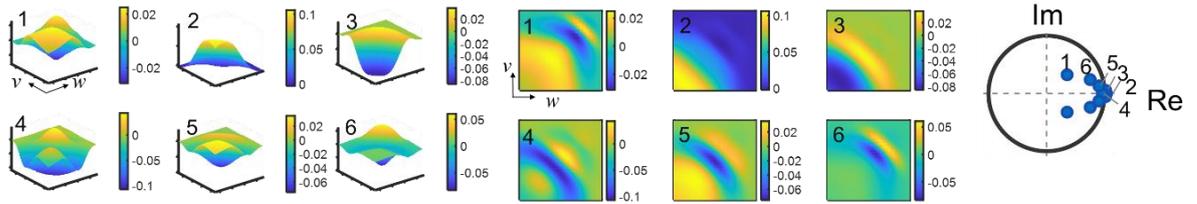

## b  Discontinuous Breakage

### Case 6

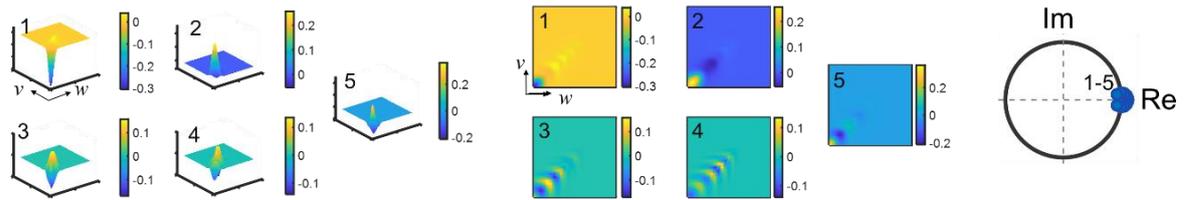

Figure 6 DMD modes $\Phi_j$ and discrete eigenvalues $\Lambda_j$ of 2D a) semi-continuous (Cases 5) and b) discontinuous (Cases 6) breakages. In Case 5, the DMD of distributions were shown from both a monodisperse and a polydisperse initial condition (IC) and Case 6 from a polydisperse IC. Top-down (2D) view of the DMD modes are shown for clearer visualization of the internal coordinate architectures. Clean data were used, sampled at 25 timepoints over the time domain [0, 5] for Case 5 and [0, 1] for Case 6. The numbering of the DMD modes and eigenvalues corresponds within each case, and the corresponding continuous-time dynamics are provided in the Appendix. The degree of truncation used in the SVD is $s = 10$, capturing > 99 % of the total energy. Note that oscillatory modes appear in complex-conjugate pairs and only one DMD mode is plotted for brevity. The colorbar represents the magnitude of mode contribution to the number density distribution.

### 3.3.3 Discontinuous Breakage

Fig. 6b presents the DMD of the distributions associated with a pure discontinuous breakage (Case 6) arising from a polydisperse initial distribution. Notably, the number density distribution exhibits distinct localization along the diagonal of the internal coordinate space (extending from top right to bottom left), which is more apparent in the 2D view. This pattern contrasts with the more spread-out peaks observed in the semi-continuous breakage case shown in Fig. 6a, where the distribution is less localized, which suggests a random breakage process in uni-direction. The localization of the distribution along the diagonal indicates fixed-size or fixed-ratio breakage, where daughter particles are consistently produced at a specific fraction of the parent size. This reflects a predictable breakage mechanism. Such pattern suggests a simultaneous reduction along both coordinate directions, indicating a splitting process that produces equal-sized daughter particles. This behavior strongly indicates that the stoichiometric kernel likely contains terms involving products of Dirac delta functions, i.e., $\delta(v-\theta_v v')\delta(w-\theta_w w')$, where $\theta_v = \theta_w = 1/2$ represent the ratios of the parent particle sizes to the daughter particle. Such stoichiometric kernels mathematically enforce the daughter particles to be fixed fractions of the parent particle size in each direction. From Fig. 6b, the clustering of discrete eigenvalues around the same spectral radius infers that the time dynamics of all modes evolve at the same frequency, suggesting a breakage rate that is independent of size, as similarly observed in Cases 1 and 3 in Fig. 5. This presents a unique scenario in which a size-independent breakage rate is coupled with a discontinuous daughter distribution, as exemplified in Case 6. In such a case, both $\boldsymbol{\Theta}_B$ and $\boldsymbol{\Theta}_D$ are simplified to a single term each which reduces the identification scenario to a full least squares problem.

**3.4 Enhanced Model Identification with DMD via Targeted Library Construction**

To demonstrate the benefits of employing DMD for improved model identification via targeted construction of the library, we evaluate the performance of mPBE-ID for two scenarios: (i) pre-DMD and (ii) post-DMD, in terms of the computation time, success rate, and the order of magnitude of the error, as illustrated in Fig. 7. Here, the bagging (mean) approach was used to aggregate the model ensemble from data bootstraps, and the computation time is the time taken to compute the solution by cb-STLS. For all pre-DMD scenarios, we imposed a library with 25 terms each for both the birth and death term libraries, i.e., $B(v'^p w'^q)$ and $D(v^p w^q)$, with exponents $p, q = -2, -1, 0, 1, 2$, yielding a total of 50 candidate terms. For the post-DMD scenarios, a refined library based on insights from DMD was used for model identification. The number of library terms used for pre-DMD vs. post-DMD is shown in Fig. 7a.

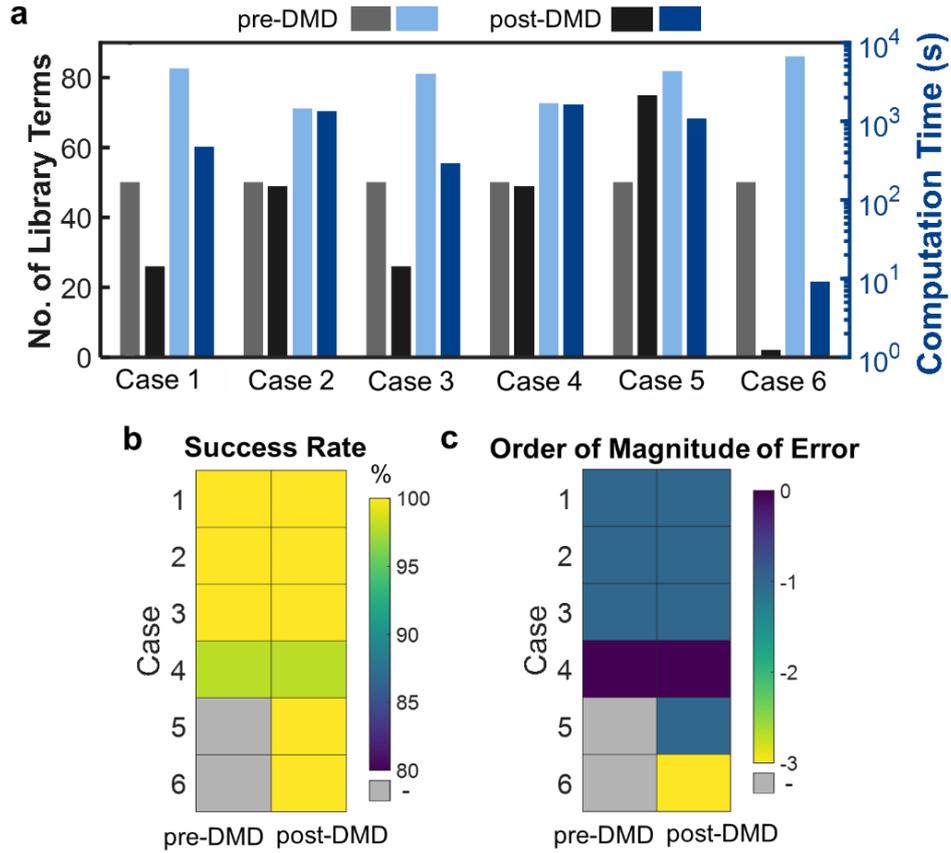

Figure 7 Comparison of pre- and post-DMD model identification performance for all cases with clean data sampled at 10 timepoints. Here, the computation time is the time taken to compute the solution by cb-STLS and the order of magnitude of error is the logarithm with base ten of the model coefficient error (Eq. (12)). The candidate library in the pre-DMD scenarios consists of 50 terms, with 25 each for both the birth and death term libraries of the forms, $B(v'^p w'^q)$ and $D(v^p w^q)$, where $p, q = -2, -1, 0, 1, 2$. The bagging (mean) approach was used to aggregate the model ensemble from data bootstraps.

For Cases 1-4, since the correct model terms are present in the library, it is not surprising that the models identified in both pre-DMD and post-DMD scenarios achieve near perfect success rates and exhibit consistently low error magnitudes below ~$O(10^0)$, as observed in Fig. 7b-c. However, without DMD insights to guide library construction, the cb-STLS computation time increases significantly, by nearly 9-fold and 13-fold for Cases 1 and 3, respectively, due to the inclusion of permutations of size-dependent rates in the death term library (Fig. 7a). This substantial computational overhead stems from the expanded library, which markedly increases the number of iterations required during STLS computation. In contrast, when size-independence is indicated by the DMD eigenvalues (Fig. 5, Section 3.3), the library is reduced by 48% in the number of terms for both cases, resulting in an approximate 90% reduction in computation time. For Cases 2 and 4, however, the computation time remains nearly unchanged, as only the size-independent rate term is excluded from the death term library. This is particularly remarkable, as substantial efficiency gains can be achieved while maintaining solution fidelity. Being able to resolve the solution swiftly yet accurately is beneficial especially for high-dimensional systems where combinatorial complexity incurs higher computational demands.

In Cases 5 and 6, as the library used in the pre-DMD scenario contains only continuous terms, the success rates and model coefficient errors could not be quantified as the correct birth terms (i.e., Dirac delta functions) were absent from the library. After distilling the key breakage patterns from DMD modes, the incorporation of a single Dirac delta function in Case 5 and a multiplicative type of Dirac delta function in Case 6, respectively, leads to accurate identification of the correct model terms with model coefficient errors below $\sim O(10^{-1})$ (Fig. 7b-c). In Case 5, as the DMD modes reveal a semi-continuous breakage pattern, all continuous terms in the birth term library were excluded in the post-DMD library and were fully replaced by permutations of single Dirac delta functions in both dimensions multiplied by $v'^p w'^q$, with $p, q = -2, -1, 0, 1, 2$. The number of birth terms thus doubles to 50, which augments the library by 50% compared to the pre-DMD scenario. Despite the larger library, the mPBE-ID captures the correct model with the order of coefficient error at $\sim O(10^{-1})$, yet reducing the computation time by $\sim$ 25% (savings of $3 \times 10^3$ s) due to reduced number of iterations during thresholding by STLS. Without prior knowledge of the potential breakage characteristics, including single

Dirac delta functions along with continuous terms would result in an excessively large library that not only increases the computational expense, but also potentially reduces the likelihood of accurately identifying the correct terms. As for Case 6, as alluded to in the preceding section, the discontinuity in the DMD modes combined with size-independent rates (Fig. 6b) refines a vast solution space to only two terms that effectively capture the correct kernels. With a mesh that progresses with a geometric ratio of two, mPBE-ID swiftly identifies the correct model as compared to the pre-DMD scenario where the continuous terms fail to capture the discontinuous breakage kernel (Fig. 7b-c). Thus, constructing a refined library informed by DMD significantly enhances model identification and reduces computational complexity, which would otherwise incur an inefficient brute force search of arbitrarily built library terms.

**3.5 The mPBE-ID is Robust Towards Noisy Data and Sparse Temporal Sampling**

As with any data-driven method, the availability of abundant and clean data is the cornerstone to its success. While increasing the resolution along the internal coordinate domain (e.g., particle size) can potentially improve model accuracy, enhancing the temporal resolution often demands substantially greater experimental resources which can limit data availability. Moreover, accurate computation of time derivatives of number density is critical for the success of mPBE-ID where noise can be amplified, especially in multi-dimensional systems. To demonstrate the capability of the mPBE-ID framework under moderate temporal sampling and noisy conditions, we evaluate the performance of three algorithmic variants: (i) without model ensembling (i.e., no bagging or bragging), referred to as the baseline algorithm, (ii) with bagging, and (iii) with bragging. All three variants were tested across an increasing number of time samples using both clean and noisy data (5% Gaussian noise), as illustrated in Fig. 8. We define regions with a success rate of $\geq 85\%$ and a model coefficient error $\leq 1$ as satisfactory. These thresholds ensure that the identified mPBE closely resembles the true structure while keeping the coefficient magnitudes within reasonable bounds. In this section, all candidate libraries were constructed based on DMD insights, as expounded in the preceding sections.

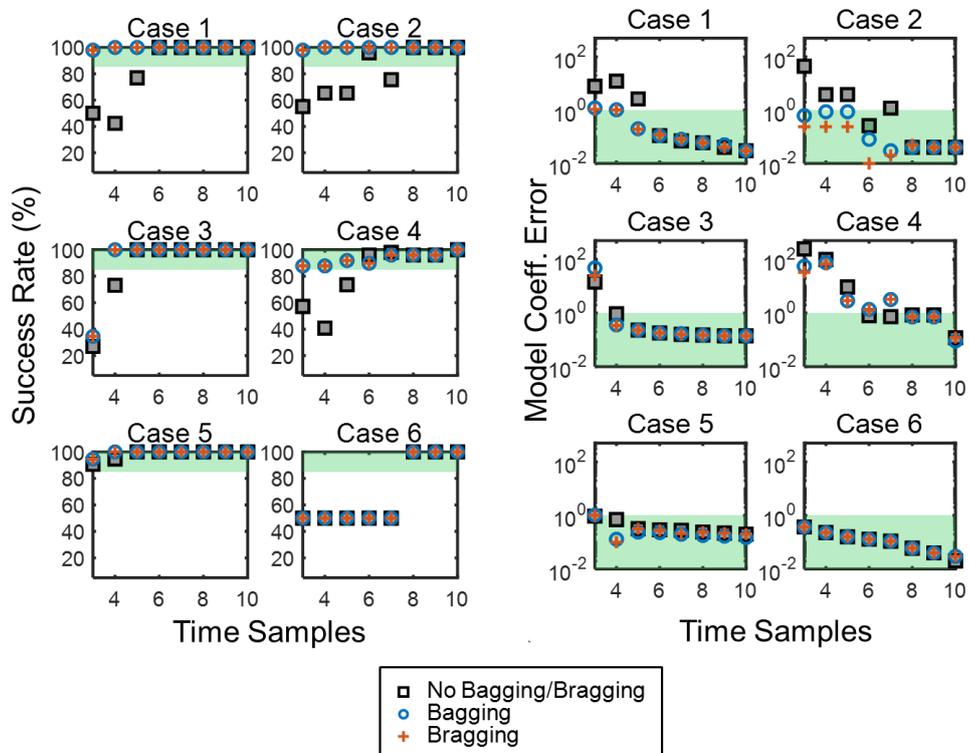

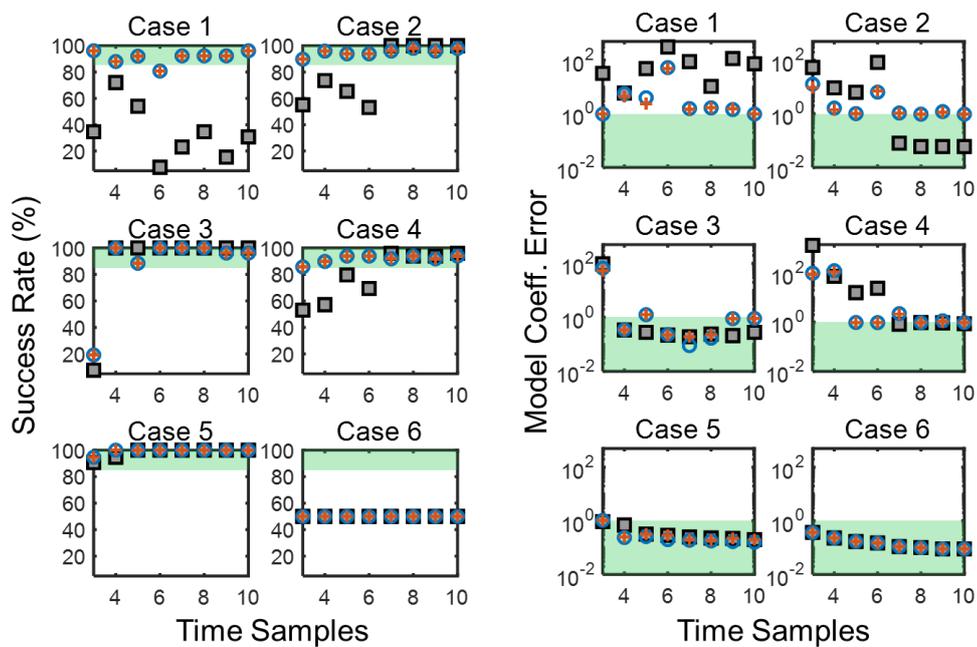

Figure 8 Model coefficient error (Eq. (12)) and success rate (%) of the models identified by mPBE-ID (with bagging and bragging) vs. a baseline framework without model ensembling

(no bagging/bragging), sampled uniformly with varying numbers of time samples across the time domain [0,5] for Cases 1-5 and [0,1] for Case 6. Results are shown for (a) clean data and (b) noisy data (5% Gaussian noise). Model ensembling via bagging and bragging was performed by aggregating predictions from 100 bootstrap replicates with replacement using the mean and median, respectively. Functional terms were retained in the aggregated model if they had an inclusion probability of $\geq 65\%$ or a coefficient of variation $\leq 1$.

For the baseline algorithm, even with high temporal sampling, we observe that noise in the data can steer the identification toward a structure that deviates significantly from the true mPBE. This behavior is particularly evident in Case 1, where the success rate drops below 40% despite high temporal sampling (> 6 timepoints) (Fig. 8b), in contrast to a 100% success rate under noise-free conditions (Fig. 8a). In Case 6, where the post-DMD library contains only two candidate terms, noise hinders the accurate recovery of the death term at high time sampling, albeit recovering the birth term correctly with high precision, thus a success rate of 50% (Fig. 8b). Overall, at all temporal sampling rates in Cases 2-5, noise has a relatively minor impact on the baseline algorithm, as success rates and model coefficient errors remain comparable to those obtained with clean data. The noise sensitivity observed in Cases 1 and 6 could plausibly be attributed to the predominantly unstable dynamical behavior of the underlying systems, as evidenced by the DMD eigenvalue spectra shown in Fig. 5 (Section 3.3). In contrast, Cases 2-5 demonstrate greater robustness to noise, potentially owing to the presence of more stable dynamic modes (Fig. 5). In reality, noise in data may be inevitable, thus our mPBE-ID employs ensembling approaches (i.e., bagging and bragging) to enhance robustness against noise-induced effects. For instance, in Case 1 (Fig. 8b), model ensembling through bagging and bragging markedly improves the success rate, elevating it into the satisfactory regime, while concurrently reducing the model coefficient error by at least an order of magnitude to values slightly exceeding $\sim O(10^0)$. Nonetheless, model ensembling can occasionally lead to slightly less accurate coefficient estimation in noisy settings. This is evident in Case 2 with more than seven time samples, where the success rate declines marginally from 100% to 96%, and the coefficient error increases by an order of magnitude compared to the baseline. Moreover, in Case 6, model ensembling does not yield improvements in success rates under conditions where noise has degraded the performance of the baseline algorithm (Fig. 8b). Despite such exceptions, in over 70% of the noise-affected scenarios, bagging and bragging lead to improved model accuracy. Further, the performance of ensemble-based models hinges on the ensembling configuration, specifically, the number of bootstrap samples and the thresholds for inclusion probability and coefficient of variation, which can be optimized to improve model accuracy.

As for the effect of the number of time samples, not surprisingly, model identification improves with more time samples in all cases with clean data, as seen in Fig. 8a. At low sampling rates (< 5 time samples), bagging and bragging enhance model identification in over 80% of scenarios. In practice, acquiring data at high temporal resolution can be resource-intensive and time-consuming. To alleviate this issue, the mPBE-ID which integrates ensembling strategies, significantly reduces model coefficient errors and improves success rates at sparse sampling rates, where the baseline algorithm often fails. Overall, mPBE-ID with bagging or bragging exhibits comparable performance in terms of structural identification. Nevertheless, bragging tends to yield more accurate model coefficients, as observed in Case 2 under clean data conditions. Therefore, users may select either bagging or bragging based on desired accuracy requirements. While mPBE-ID may occasionally fail to reduce coefficient errors to satisfactory levels at low sampling rates (e.g. Case 4), both bagging and bragging consistently achieve high success rates, suggesting that the correct model terms are reliably identified albeit with inaccurate coefficient values. This is especially valuable given that knowledge of the correct terms can be used to revise the library or guide further experimental data acquisition to improve the model.

## 4.0 Conclusion

We have developed a data-driven pipeline for discovering multi-dimensional breakage population balance equations. Our Multi-Dimensional Breakage Population Balance Equation Identification (mPBE-ID) framework builds upon the sparse regression technique to uncover sparse, interpretable and physically realizable mPBEs directly from data by selecting relevant terms in the dynamics from a library of candidate functions. Being the first framework to enable the discovery of multi-dimensional breakage population balance equations directly from data, the mPBE-ID introduces various strategies to uncover mPBEs in an efficient manner: (i) imposition of breakage-informed constraints within the least squares algorithm, (ii) targeted construction of library candidate functions via insights from DMD, and (iii) robust handling of noisy and low temporal resolution data through model ensembling (bagging/bragging). As a result, the mPBE-ID efficiently identifies mPBEs for various breakage scenarios and is reasonably robust against noisy and limited data situations.

We anticipate that the mPBE-ID will serve as a foundation for systematic discovery of mPBEs, which in turn will lead to wider adoption of multi-dimensional population balances in emerging fields such as crystallization engineering, atmospheric science, and pharmaceutical manufacturing. As part of our effort to address the long-standing challenge in general PBE identification, future work should extend the framework to other particulate phenomena such as multi-dimensional aggregation, growth and nucleation. In addition, it is of immense importance to also look into breakage scenarios where the stoichiometric and rate kernels cannot be readily decomposed into a linear combination of candidate terms, e.g., those which involve more sophisticated functions such as the multiplication of an exponential function with another function, as they commonly appear in fields involving droplet/bubble breakup [47]. Moreover, imposing physical constraints on the stoichiometric kernel within the mPBE-ID framework may be valuable for ensuring consistency with breakage heuristics, such as conservation of moments according to the number of daughter particles generated through breakage. While this work affirms the theoretical potential of mPBE-ID to discover mPBEs from controlled synthetic datasets where the true solutions are known, the application of mPBE-ID to real experimental data, which requires exhaustive deliberation on the target process, will be communicated in our forthcoming work.

**Acknowledgement**

We would like to acknowledge the Graduate Research Excellence Scholarship from Monash University Malaysia for the financial support to Suet Lin Leong.

**Appendix**

**A1. Continuous-Time Evolution of DMD Modes**

The continuous-time evolutions of the DMD modes shown in Figs. 5-6 (Section 3.3) are presented here, with the numbering in Fig. A.1 corresponding to those in Figs. 5-6 for each case. As seen in Eq. (9), the expression $\exp(\Omega_j t)$ describes the temporal evolution of each DMD mode, where the real part of the continuous-time eigenvalue $\Omega_j$ (i.e., $\text{Re}(\Omega_j)$) determines growth or decay in a DMD mode, while the imaginary part $\text{Im}(\Omega_j)$ determines the oscillatory frequency. Fig. A.1 plots the real part of $\exp(\Omega_j t)$, i.e., $\text{Re}\left[\exp(\Omega_j t)\right]$, for all cases studied. From the figure, it can be observed that DMD modes with unstable eigenvalues (discrete eigenvalues outside unit circle in Fig. 5-6) have continuous-time dynamics that grow with time (e.g., Modes 1-4 of Case 1), and vice versa for stable modes (e.g., Modes 5-6 of Case 1).

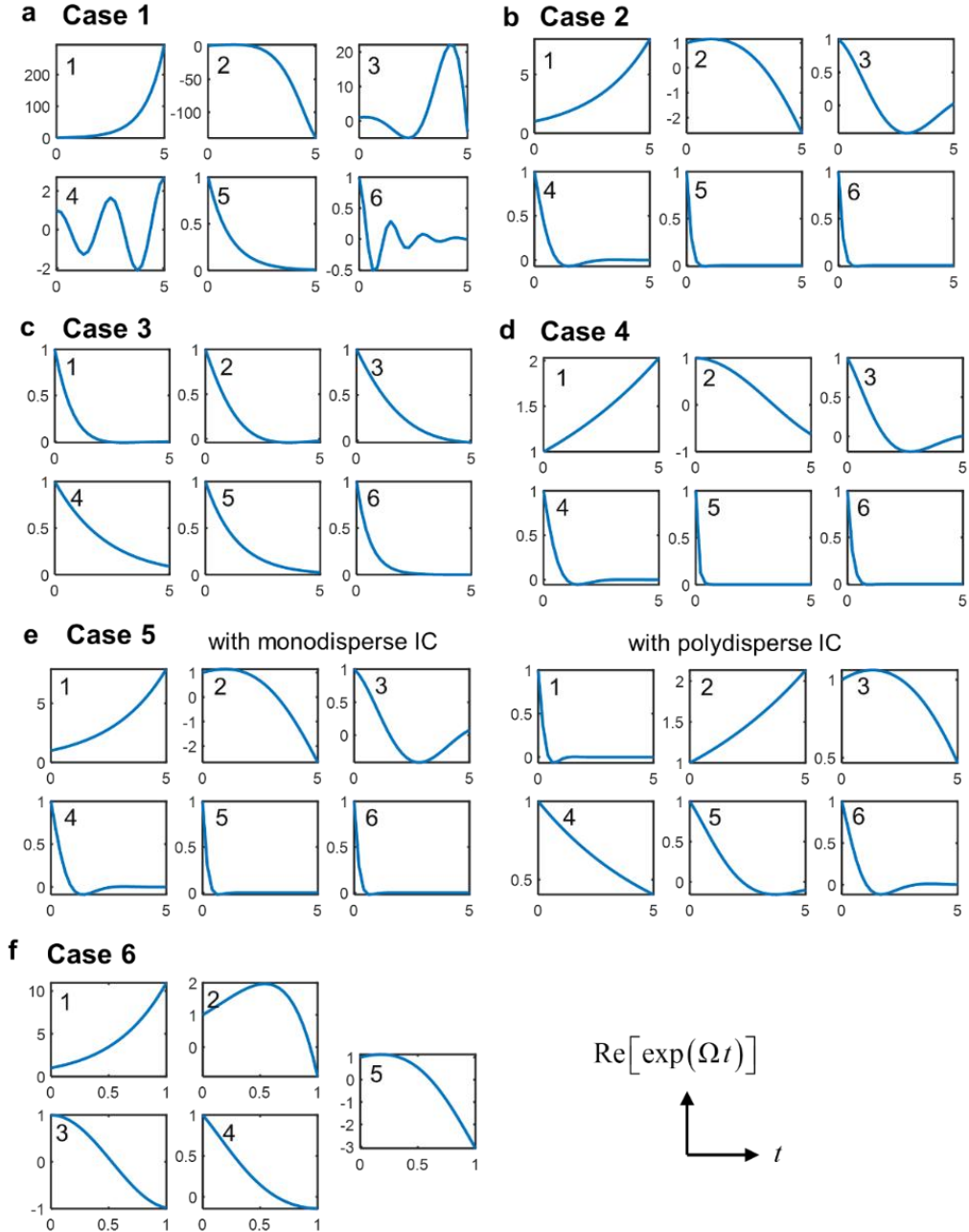

Figure A.1 Continuous-time evolution of the DMD modes for all cases, where a growing trend indicates amplification and a decaying trend indicates damping of the corresponding mode. The plots represent the real part of the time-evolving dynamic modes, $\text{Re}\left[\exp(\Omega_j t)\right]$, where $\Omega_j = \ln \Lambda_j / \Delta t$ is the $j$-th continuous-time eigenvalue and $\Lambda_j$ is the $j$-th discrete-time eigenvalue. Note that negative values of $\text{Re}\left[\exp(\Omega_j t)\right]$ should be interpreted together with the corresponding signs of the DMD mode $\Phi_j$ magnitudes (Figs. 5-6) to assess the overall contribution of each mode to the system dynamics (Eq. (9)). For instance, a DMD mode with negative magnitudes coupled with a negative temporal eigenvalue will lead to an overall positive contribution to the dynamics. For example, in Mode 2 of Case 1 (Fig. 5), the largest negative magnitude in the DMD mode are concentrated in the smaller-sized particles. This, when coupled with a negative temporal eigenvalue (this figure), indicates that the small-sized particles are in fact growing.